\documentstyle[psfig]{aa}
\begin{document}
\thesaurus{ 03.20.2; 11.04.1; 11.11.1; 11.17.1; 11.17.4 3C196 }

\title {Cold Gas Kinematics in an $L_*$ Spiral Galaxy at $z=0.437$:
The Nature of Damped Lyman-$\alpha$ Absorbers}

\author{F.H. Briggs\inst{1}, A.G. de Bruyn\inst{1}\fnmsep\inst{2},
R.C. Vermeulen\inst{2} }
\institute{Kapteyn Astronomical Institute, P.O. Box 800, 9700 AV
  Groningen, The Netherlands\\
\and ASTRON, P.O. Box 2, 7990 AA, Dwingeloo, The Netherlands
}

\offprints{ }
\date{Received ~; accepted ~}

\maketitle

\markboth{ Kinematics of a $z-0.44$ Spiral Galaxy }
{ Kinematics of a $z-0.44$ Spiral Galaxy  }

\begin{abstract}  
 
 Westerbork Radio Synthesis Telescope observations
of the redshifted 21cm line absorber against the $z_{em} = 0.871$
double lobed quasar 3C196 show that  the
intervening absorber is 
an $L\approx L_*$ spiral galaxy (3C196-G1)
and that the absorbing layer of cold
gas extends to radii of at least 30~$h_{50}^{-1}$kpc.  The new data solve
several long standing puzzles about this system by (1) discovering
a second 21cm absorption feature, corresponding to  absorption
against the NE lobe of the background radio source and (2) spatially
``resolving'' the two absorption features to isolate the absorption 
along the two lines of sight to the opposing radio lobes. These findings
resolve the disagreement in redshift between the UV metal and 21cm lines,
and as well as demonstrating that the neutral layer does absorb both
lobes of the background radio source.
Simple kinematic models with an inclined, rotating gas disk
match the observed 21cm profile and are also compatible with both
the redshift and velocity spread of the absorption measured in UV
resonance lines along a third, independent line of sight to the
quasar nucleus and  with the lack of 21cm  absorption in as
earlier VLBI experiment that was sensitive to opacity against the
hot spot in the northern lobe.
The inferred rotation speed and luminosity for the galaxy are compatible
with the $z\approx 0$ Tully-Fisher Relation.

This system illustrates well how 21cm absorption against extended background
radio sources is a powerful tool in determining the nature of the damped
Lyman-$\alpha$ class of QSO absorption line system.

\end {abstract}


\section{Introduction}

Cold neutral gas traces gravitational potential wells, and in the
nearby universe, astronomers use the 21cm line as a dynamical indicator to
obtain 
the total amount and distribution of gravitating mass associated
with galaxies.  At high redshift, the neutral gas mass constitutes a larger
fraction of the observable luminous mass in baryons
(Lanzetta et al 1995, Storrie-Lombardi \& Wolfe 2000), making
the HI itself a more important gravitating element, and
the stars less so.  This increases the importance of selecting
galaxies at high redshift based purely on their neutral gas content,
since these objects are likely to point to deep gravitational
potentials where the progenitors of the
galaxies of the present epoch are  forming.

Despite presenting different opinions on the nature of the
high HI column density QSO absorption line systems with
$N_{HI} \geq 2{\times}10^{20}$ atoms~cm$^{-2}$ known as ``damped Lyman-$\alpha$ (DLa),
both Prochaska \& Wolfe (1998) and Haehnelt et al (1998) do agree that
the DLa systems are ``the progenitors of the present day normal galaxies.''
Since the 21cm line requires column densities of HI in excess of
${\sim}10^{20}$ atoms~cm$^{-2}$ to create strong absorption lines, the
redshifted 21cm absorbers are identified with the DLa
class of quasar absorption-line and thus with evolving galactic potentials.

Due to its early identification by Brown \& Mitchell (1983), the
21cm line observed at $z=0.437$ against the extended radio quasar
3C196 ($z_{em}= 0.871$) has received extensive follow-up with a variety
of techniques, including ground-base and HST imaging and spectroscopy.
The 3C196 line of sight has
received additional attention, due to its association with DLa's and
its low redshift, which makes detection of a possible optical counterpart
to the absorber and the study of the environment of the absorber
easier than for the higher $z$ DLa's that are selected most readily
in ground-based spectroscopy of the Lyman-$\alpha$ line at $z>1.7$.
The absorption redshift was subsequently selected in a survey for
MgII absorption  against steep spectrum quasars by Aldcroft et al (1993, 1994).

The optical identification of two candidates (3C196-G1 and
3C196-G2) for an intervening
absorbing galaxy were made by Boiss\'{e} \& Boulade (1990).
Subsequent HST imaging has shown that G1 is a large barred
spiral located about 1.5$''$ south-east of the quasar (Cohen et al 1996,
Le Brun et al, 1997, Ridgway \& Stockton 1997). The outlying spiral
arms of G1 appeared to extend across the two extended radio lobes
of the background radio quasar, 
leading to the puzzle posed by Cohen et al over why only one relatively
narrow absorption profile was observed in the 21cm line spectrum,
when a velocity spread of several hundred km~s$^{-1}$ would be 
expected  for absorption over the full extent of the G1 galaxy.
G2 is due north of the quasar nucleus.

A second puzzle arose from the spectroscopic studies.  Ground based
spectroscopy (Foltz et al 1988) showed that the metal-line absorption
was spread over a wider range than the 21cm absorption, but that
the redshift of the center of the metal line profiles fell at
${\sim}50$~km~s$^{-1}$ greater than   the 21cm line redshift (Cohen
et al 1996).

A third important constraint comes from the spectral-line VLBI
observation of Brown et al (1988), which detected no absorption
against a marginally resolved hot spot in the NE radio continuum
lobe and implied that the absorption must take place in a spatially
extended absorber against the more diffuse and extended radio
continuum structure, thus placing a lower limit on the size of
the absorber of $13h_{50}^{-1}$kpc (Foltz et al 1998).

In summary, prior to the observations we report here, there was a candidate
galaxy for which no redshift had been measured to confirm its role
as the site of the 21cm absorption, and there were several
puzzles about how the galaxy's kinematics might explain
both the 21cm absorption profile and the line widths and redshift of the metal
lines.
These new 21cm line observations resolve these ambiguities, establishing
the luminous barred spiral 3C196-G1 as the absorber, as well as
providing a measure of its dynamical mass. The system remains an interesting
candidate for high-spatial resolution radio aperture synthesis, and
provides a view for how these techniques could be applied to 
the highest redshift DLa absorbers in order to determine their 
nature at much higher redshifts.

\section{The WSRT Observations}
 
\begin{figure}
  \psfig{figure=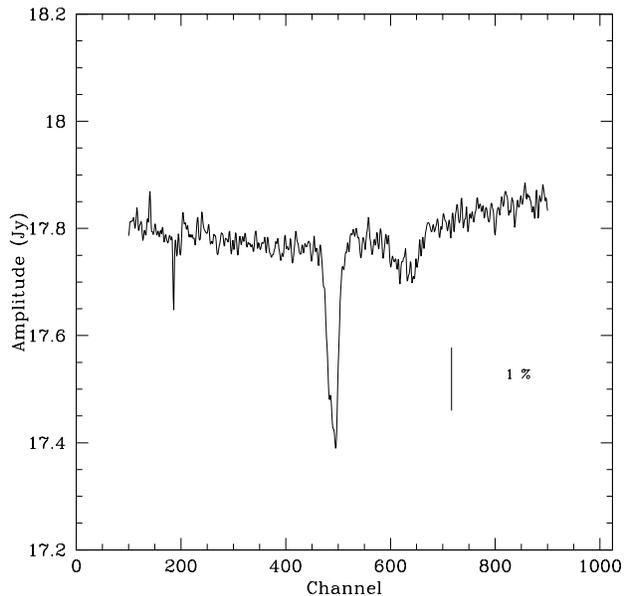,width=8.5cm,angle=0}
  \caption[]{Compound interferometer
spectrum of 3C196. The total spectral range displayed measures
about 4 MHz or 1200 km/sec. The velocity resolution is about 4.5
km~sec$^{-1}$.  }
  \label{ci.fig}
\end{figure}

New Westerbork Synthesis Radio Telescope
observations were conducted in the Compound Interferometer
(CI) as well as in regular synthesis mode. In the CI-mode we
crosscorrelate the signal from two tied-arrays each representing
the summed signal of 6 telescopes (Chengalur et al 1996).
In doing so we can use the
full correlator capacity to record both a wide band (5 MHz) and
attain high spectral resolution (15 kHz). A 12 hour observation
on December 21 1996 was distributed equally between 3C196 and the
bandpass calibrator source 3C147 with a dutycycle of 2 hours. The
resulting spectrum is shown in Fig.~\ref{ci.fig}. The 2\% deep absorption
line discovered by Brown \& Mitchell is seen around channel 500.
However, we also detected a second absorption line. This line,
visible near channel 650 is both wider (about 100 km/sec) and
shallower (0.3\% peak absorption) [The sharp feature near channel
200 is due to interference]. No baseline was removed from the
spectrum.

 A second observation was done on January 6 1997 in regular
synthesis mode using the line backend (Bos et al 1981) with 2.5 MHz and 64
channels with the central frequency halfway between the two
absorption features. A total of 10 hours of integration was spent
on 3C196 with the available 13 telescopes. The calibrator 3C147
was observed for a total of 3 hours before and midway during the
3C196 observation. Following a selfcalibration on 3C147 we
transferred the complex passbands, on a telescope by telescope
basis, to the visibilities of 3C196. The data were then Fourier
transformed to yield a cube of 56 images (channels 3-58). 3C196
is unresolved by the WSRT synthesized beam at 988 MHz, which has half-power
full-width of 20$''{\times}26''$, and we therefore determined the peak
amplitude of the source as a function of frequency. The resulting
spectrum is shown in Fig.2. No baseline had to be removed. The
quality of the spectrum is indeed excellent. Both absorption
features are seen with high S/N and there is no doubt about the
reality of the shallower feature at 988.0 MHz. We note this
because this feature was not seen in the spectrum reported by
Brown \& Mitchell. The superior baseline stability of
interferometers compared to those obtained using a single dish is
obvious. The passband is stable to better than one part in a
thousand and is in fact limited by thermal noise in the 3C196 and
3C147 data.

The redshift measured for the optical and UV lines (Foltz et al 1988)
falls between the two 21cm absorption features.

\begin{figure}
  \psfig{figure=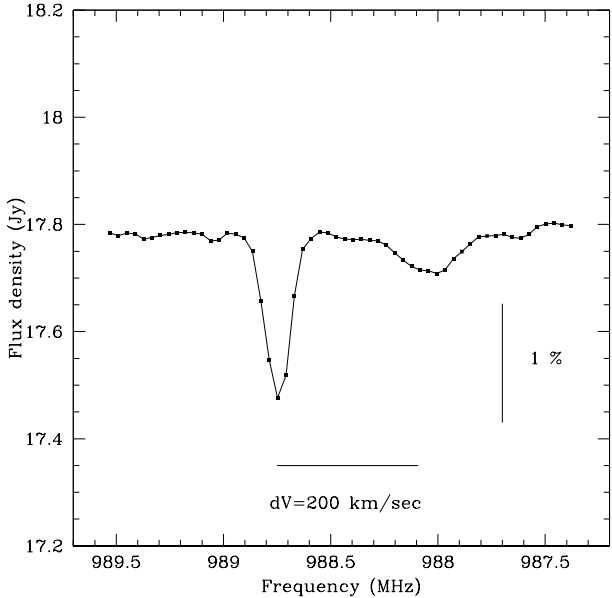,width=8.5cm,angle=0}
  \caption[]{Regular synthesis mode
spectrum of 3C196. The spectral resolution is about 24 km~sec$^{-1}$.  }
  \label{dxb.fig}
\end{figure}

 The   WSRT synthesized beam at this frequency is ${\sim}25''$,
and this is much larger than the size of the radio source.
However, the spectral dynamic range of the data, in amplitude as
well as in phase, is so good ($\sim$4000:1) that we should be able to determine
the centroid of the source as a function of frequency to 1 part
in 8000 of the synthesized beamwidth. This is done with the
NEWSTAR program NMODEL that fits a model to the UV data. Fig. 3
shows the derived source centroid position as a function of
frequency. Note that the angular scale is measured in
milli-arcseconds. In three groups of continuum channels outside
the absorption lines, the centroid position is found to be
independent of frequency to within a few mas, as could be
expected on the basis of the S/N ratio. This centroid lies about
halfway between the two radiolobes. However, in the absorption
channels we see a systematic shift of the source centroid. In the
channels corresponding to the deepest absorption feature the
source centroid shifts by about 0.040$''$ to the north-east
with the largest shift occurring in the channels where the
absorption line reaches its lowest value of about -1.6\%.
Correcting the observed shift by the channel dependent depth of
the absorption line, we can deduce the true location of the deep
absorption feature to lie about 2.5$''\approx 0.040/0.016$ in the opposite
direction, i.e. toward the south-west. This is precisely the
location of the brightest hotspot relative to the continuum
centroid as deduced from published 408 and 1666 MHz MERLIN images
(Lonsdale \& Morison 1983). The true depth
of the absorption line against that hotspot must be about 4\%
since this hotspot is responsible for nearly 50\% of the total
flux density at 988 MHz.

\begin{figure}
  \psfig{figure=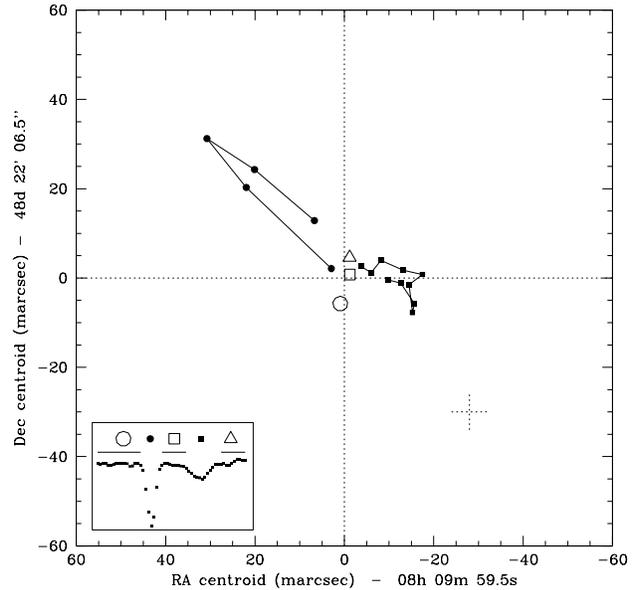,width=8.5cm,angle=0}
  \caption[]{ Centroid position of 3C196 as a function of
frequency. Note that the scale is in milli-arcseconds. For the
deep and shallow absorption features five, respectively, ten
channels are shown. The code for the frequency is shown at the
lower left. The 1 sigma error bar per channel (4 mas) is shown by
the dotted cross. The symbols for the three continuum ranges are
averages over about 17, 10 and 11 channels respectively, and have
a much smaller error.}
  \label{centroid.fig}
\end{figure}

The shallow absorption feature, on the other hand, shows an
average centroid shift of about 0.015$''$ in the opposite
direction implying that the absorbed part of the radio source
lies about 3.7$''\approx .015/.04$ east-north-east of the continuum centroid.
This corresponds to the location of the eastern lobe of 3C196.
The larger velocity width of this line suggests that gas with a
range of velocities is present in front of this extended lobe.
Since the total flux in this lobe (at 988 MHz) is about 15-20\% of
the total continuum flux the optical depth must be at least 2\%
and probably several times larger since gas at a single velocity
will be covered by only part of the lobe. 
 The location of the absorbing gas leaves no doubt that the
barred spiral seen in the HST image is responsible for the 21cm
absorption.

\section{Modeling}
 
\begin{figure}
  \psfig{figure=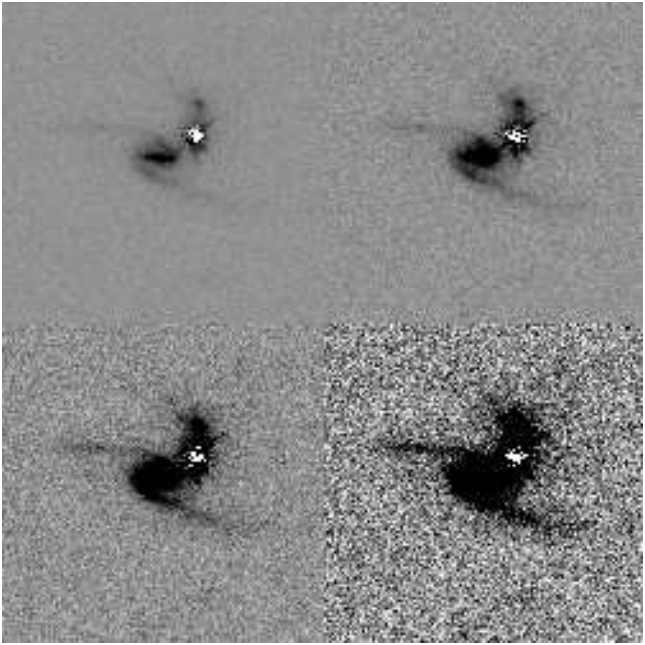,width=8.5cm,angle=0}
  \caption[]{ HST image of 3C196 after PSF subtraction
(Ridgway \& Stockton 1997) shown for four contrasts.}
  \label{hst.fig}
\end{figure}

A simple kinematic model consisting of a differentially rotating
flat disk of neutral gas can explain the observational results
of Figs.~1-3.  We make use of the HST image in Fig.~\ref{hst.fig}
 (kindly provided by
S. Ridgway) to locate the center and approximate orientation of the
optically emitting galaxy, thereby imposing additional constraints on the
geometry of the model. The UV resonance line spectroscopy against
the QSO nucleus, both from ground-based telescopes and from HST,
provides kinematical information along an additional line of sight
through the galaxy. For this illustrative model, we assume a uniform
gas distribution throughout the disk; the HST image shows a barred galaxy
with prominent arms, which is unlikely to provide uniform optical depth.
The data at hand does not justify a more detailed model.

The modeling constrains the range of galaxy model parameters but
does not produce a unique fit. In exploring the parameter space,
we have adjusted the following eight parameters:
the offset of the galaxy center $\Delta RA$ (measured eastward) and
$\Delta Dec$ with respect to the origin defined in the 1666 MHz
maps of Lonsdale \& Morison (1983),
$R_{\rm 21}$ radius of the gaseous absorber, $i$ inclination of the rotating disk,
$pa$ position angle of the receding side of 
disk major axis, $V_{\rm rot}$ rotation speed for
a flat rotation curve, $\tau_{\rm 21}$
21cm line optical depth measured perpendicular
to the disk, and $\sigma_{\rm 21}$
velocity dispersion of the gas. The
parameters for four examples are listed in Table~\ref{model.tab}, along
with derived quantities describing $N_{\rm los}(HI)$ the
line of sight neutral hydrogen column density
($N_{\rm los}(HI)=1.82{\times}10^{18}\sigma_{\rm 21}\tau_{\rm 21}T_{\rm s}/\cos i$,
where a spin temperature $T_{\rm s} = 100$~K is adopted for these estimates),
inferred $M_{\rm HI}$ neutral hydrogen
mass within $R_{\rm 21}$,
$M_{\rm dyn}=V_{\rm rot}^2R_{\rm 21}G^{-1}$ the dynamical mass
and the $M_{\rm dyn}/L_{\rm B}$ ratio of dynamical
mass to optical B-band luminosity. The derived parameters are discussed in
more detail in Sect. 4.

\begin{table*}
\caption{Uniform Disk Models}
\begin{tabular}{lrccrrcrcrccrrc}
\hline \hline
 & & & & & & & & & &  \\
\multicolumn{2}{c}{Model} &  
\multicolumn{2}{c}{ Center } & 
\multicolumn{1}{c}{p.a.} & 
\multicolumn{2}{c}{Radius} &
\multicolumn{1}{c}{Incl.} &
\multicolumn{1}{c}{$V_{\rm rot}$} &
\multicolumn{1}{c}{$\tau_{\rm 21}$} &
\multicolumn{1}{c}{$\sigma_{\rm 21}$} &
\multicolumn{1}{c}{$N_{\rm los}(HI)$} &
\multicolumn{1}{c}{$M_{\rm HI}$} &
\multicolumn{1}{c}{$M_{\rm dyn}$} &
\multicolumn{1}{c}{$\frac{M_{\rm dyn}}{L_{\rm B}}$}
\\
                   &       & 
\multicolumn{1}{c}{$\Delta RA$} & \multicolumn{1}{c}{$\Delta Dec$} &    
\multicolumn{1}{c}{ $pa$}&
\multicolumn{2}{c}{$R_{\rm 21}$} &
\multicolumn{1}{c}{$i$} &  & & & 
\multicolumn{2}{c}{ ( $T_{\rm s}=100$ )} & \\
                   &       &  
\multicolumn{1}{c}{[$''$]} & \multicolumn{1}{c}{[$''$]} & 
\multicolumn{1}{c}{[deg]}  &
\multicolumn{1}{c}{[$''$]} & \multicolumn{1}{c}{[kpc]} & 
\multicolumn{1}{c}{[deg]}  &
\multicolumn{1}{c}{[km s$^{-1}$]} & 
                           &
\multicolumn{1}{c}{[km s$^{-1}$]} &
\multicolumn{1}{c}{[cm$^{-2}$]} &
\multicolumn{1}{c}{$10^9M_{\odot}$} &  
\multicolumn{1}{c}{$10^{11}M_{\odot}$} &  
\multicolumn{1}{c}{$\frac{M_{\odot}}{L_{\odot}}$ }  
\\
\hline \\
I  &fast  & 2.8& 1.5& -80& 7.0  & 52&70& 250& 0.06& 7& 5.6(10$^{20}$) 
&13.3 & 7.6 & 14\\  
II & Cohen& 2.8& 1.5& -85& 5.5& 41&63& 250& 0.03& 7& 2.1(10$^{20}$) 
& 4.0 & 6.0 & 11\\  
III& slow& 2.8& 1.5& -95& 5.5& 41&70& 180& 0.06& 7& 5.6(10$^{20}$) 
& 8.2 & 3.1 & 6\\ 
IV &slow & 2.8& 1.5& -100& 5.0& 37&70& 180& 0.06& 7& 5.6(10$^{20}$) 
& 6.7 & 2.8 & 5\\ 
 & & & & & & & & & &  \\
\hline\hline \\
\end{tabular}
\label{model.tab}
\end{table*}

\begin{figure}
  \psfig{figure=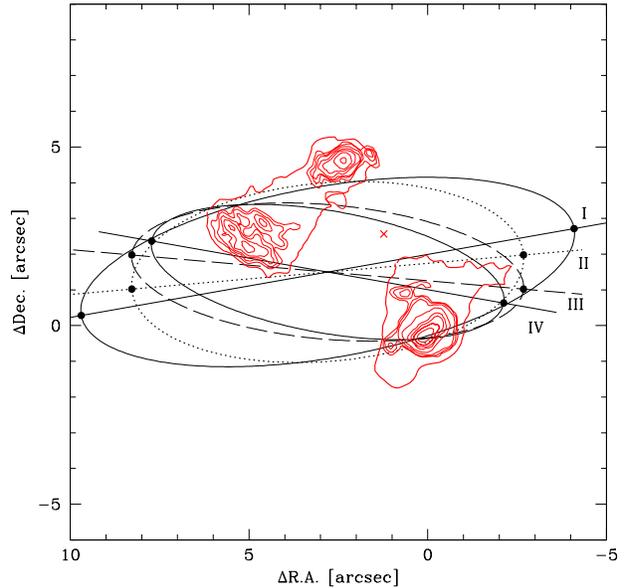,width=8.5cm,angle=0}
  \caption[]{Orientations and sizes of the four disk models overlaid
on radio continuum contours.  The representative models are 
labeled (I, II, III, IV) next to their major axes. 
The outer radio contour is taken from 
the 5 GHz map of Oren as shown by Cohen et al (1996); the higher contours
are 1666 MHz MERLIN data from Lonsdale \& Morison (1983).  The position of the
optical quasar and radio nucleus is marked with an ${\times}$.}
  \label{ovals.fig}
\end{figure}

\begin{figure}
  \psfig{figure=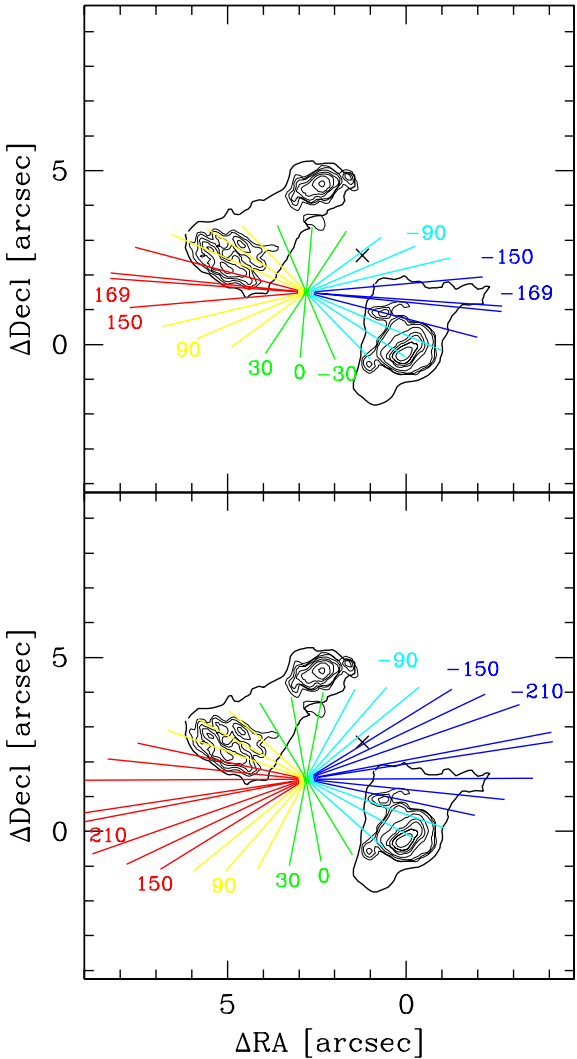,width=9.2cm,angle=0}
  \caption[]{The model velocity fields for models I (lower) and III (upper)
overlaid on the radio continuum contours from Fig.~\ref{ovals.fig}.}
  \label{velfields.fig}
\end{figure}

The sizes and orientations of the absorbing disk models are illustrated
in Fig.~\ref{ovals.fig} by overlaying the outer boundaries of the disks
on the contours of radio brightness. There is overall excellent
agreement among the radio continuum maps for 3C196,
 including the 408 and 1666 MHz MERLIN maps of Lonsdale \& Morison (1983),
the 5 GHz maps of Brown et al (1986), Lonsdale (1984), and
Oren (1996, shown in Cohen et al 1996). Here, we have chosen to illustrate
the structure with the 1666 MHz contours, which are the closest in
$\log\;\nu$ to the redshifted line frequency at 988~MHz. Since the MERLIN  maps
lack surface brightness sensitivity for weak extended emission, we have
drawn as the lowest contour the outermost VLA 5 GHz contour of Oren,
which has comparable resolution to the 1666 MERLIN map. The bulk of
the flux density is contained in the 1666 MHz structure, so
the distribution of source intensity used for computing model absorption profiles uses purely the 1666 MHz MERLIN map. 
The origin of the $x$-$y$ coordinate system (aligned EW and NS on the sky)
is located at the S-W hot spot in the S-W  radio continuum lobe.
In this system, the quasar nucleus is located at $+1.23'',+2.56''$.

Fig.~\ref{velfields.fig}
illustrates the velocity fields for the flat rotation curve models
I and III from Table~\ref{model.tab}.  The absorption spectra $S(V)$ are
computed by integrating the continuum
source intensity, $I({\bf r})=I(x,y)$, attenuated
by the velocity dependent absorption, ${\exp}(-\tau_{\rm 21}(V,{\bf r})/{\cos}\;i)$,
over the extent of the continuum source:
$$
S(V) =
\int_{{\rm all\;} {\bf r}}e^{-\tau_{\rm 21}(V,{\bf r})/{\cos}\;i}I({\bf r})\; d{\bf r}
$$
Here, {\bf r} is the position vector measured in the plane of
the sky relative to the $x$-$y$ coordinate system defined in 
Fig.~\ref{ovals.fig}: ${\bf r}=x\hat{\bf i}+y\hat{\bf j}$.

The source centroid ${\bf r}_{\rm c}(V)$ as a function of velocity results
from
$$
{\bf r}_{\rm c}(V) = \left[ S(V) \right ]^{-1}
\int_{{\rm all\;} {\bf r}}{\bf r}\;e^{-\tau_{\rm 21}(V,{\bf r})/{\cos}\;i}I({\bf r})\; d{\bf r}
$$

\begin{figure}
  \psfig{figure=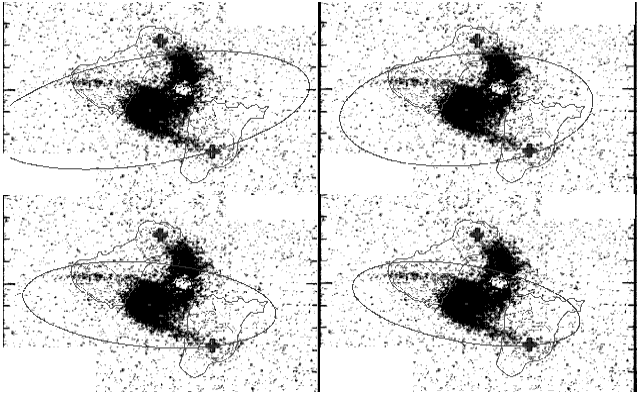,width=8.5cm,angle=0}
  \caption[]{ The ovals representing the four models from Fig.~\ref{ovals.fig}
overlaid on the HST image from Ridgway \& Stockton (1997). Ovals are
centered on galaxy 3C166-G1. The stellar image of the quasar nucleus
has been subtracted by Ridgway \& Stockton. Upper
row shows models I (left) and II (right); lower row has III (left) and
IV (right).  Crosses indicate the locations of the hot spots in the radio
lobes.  }
  \label{overlays.fig}
\end{figure}

The HST and ground-based optical imaging provides some constraints on
the size, location and orientation of the disk model. Here, we begin
by centering the absorbing disk on the galaxy tentatively identified
by Boiss\'{e} \& Boulade (1990), Cohen et al (1996) and Le Brun et al
(1997), and we conclude that the model satisfactorily explains the 
double-featured, 21cm line profile (in accord with the prediction of 
Cohen et al) as well as the velocity dependent centroid shift
shown in Fig.~\ref{centroid.fig}. In the $x$-$y$ coordinate system adopted here,
the galaxy center is located at $+2.8'', +1.5''$.  All models in
Table~\ref{model.tab} have orientations and extents that encompass the
full extent of the spiral arms.  Fig.~\ref{overlays.fig} shows the
4 ovals defined by models I-IV overlaid on the HST image of 
Ridgway \& Stockton (1997); we chose this image for comparison, due to
its high contrast and ease of access through the online ADS database.
Ridgway \& Stockton have subtracted a stellar point-spread function to
remove the bright QSO nucleus, and they discuss the possible connection between
some of the remaining nebulosity and the QSO itself at $z_{em}=0.871$. 
Here, we concentrate on the nature of the intervening spiral galaxy
3C196-G1 to the south-east of the QSO nucleus.
 
\begin{figure}
  \psfig{figure=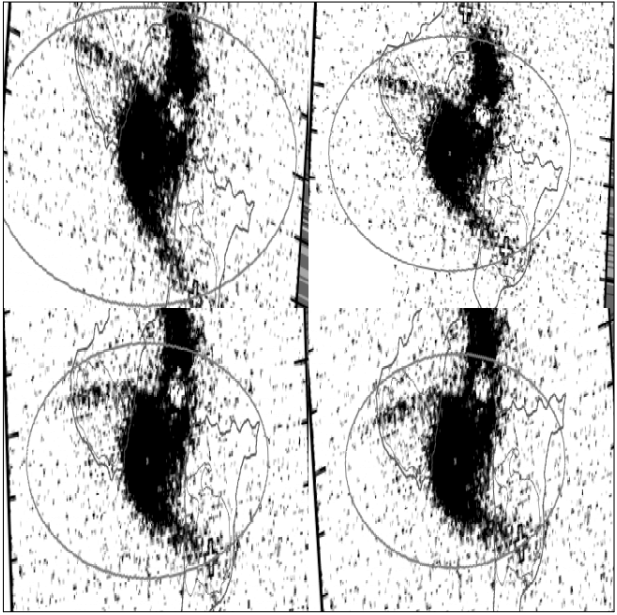,width=8.5cm,angle=0}
  \caption[]{Deprojected views of the optical surface
brightness. Optical image of Ridgway \& Stockton is projected
to a face-on view, using the orientation parameters determined for 
the kinematical models. Ovals from Fig.~\ref{overlays.fig}
become circular in this view. Upper
row shows models I (left) and II (right); lower row has III (left) and
IV (right).}
  \label{deprojoverlays.fig}
\end{figure}

Since the models define a disk orientation, it is straightforward to 
deproject the galaxy images to view the galaxy as though it were 
face-on. The four images, deprojected using the model
parameters in Table~\ref{model.tab}, are shown in Fig.~\ref{deprojoverlays.fig}.

\begin{figure}
  \psfig{figure=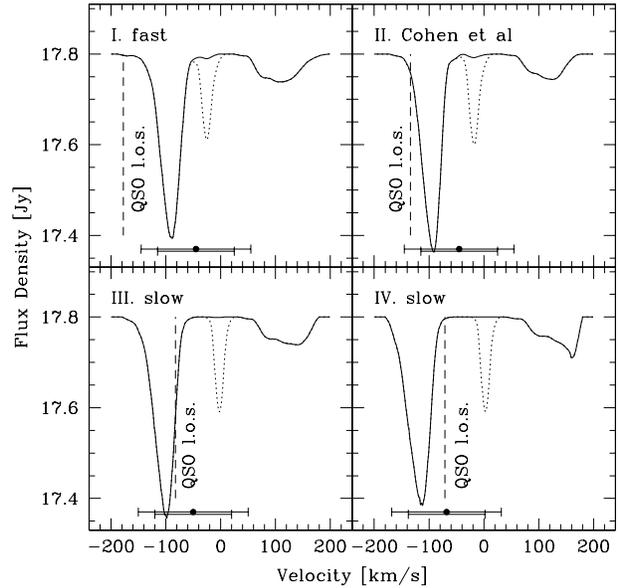,width=8.5cm,angle=0}
  \caption[]{Model spectra for the four representative models
shown in Figs.~\ref{ovals.fig}-\ref{deprojoverlays.fig}. The
model spectra are to be compared with the observed spectrum
shown in Figs.~(1\&2).
Parameters for the models are summarized in Table~1.
Velocities are specified relative to the galaxy systemic
velocity.  The light dashed curve indicating a narrow
absorption feature near the center of the profile
demonstrates how the spectrum would be influenced by
a disk that extends north to cover the northern lobe hotspot.
The vertical dashed line represents the velocity
predicted by the model for the line of sight to the quasar 
nucleus.  The error bars indicate the FWMH velocity range
of the UV/optical lines, measured for the MgII, FeII lines (wider range) 
and the MgI, CaII lines (narrower range); the resonance lines 
are positioned 50 km~s$^{-1}$ higher in redshift than the stronger
component of 21cm line absorption (Cohen et al 1996).
}
  \label{spectra.fig}
\end{figure}

The model spectra for the four models are shown in Fig.~\ref{spectra.fig}.
The double-featured absorption profile is a natural consequence of the
absorbing disk covering two distinct lobes of the background continuum
source. In these models, the shapes of the two individual
profiles depend  on the distribution of continuum intensity behind the
velocity gradient of the differentially rotating, disk absorption.
No variation in the 21cm line optical depth has been put in the model,
except that the disk is limited in radius to $R_{\rm 21}$.
Models I, II and III provide good recovery of the absorption lines
of Figs.~1-2. 
Model  IV shows an example of a model that begins to
fail due  to  the appearance of a distinct
narrow feature in the low-redshift component due to the placement of
the major axis (a region of the velocity field with low velocity gradient)
over the diffuse
eastern continuum lobe, leading to a feature that is akin to the horn of a inclined spiral
galaxy emission profile.

Fig.~\ref{spectra.fig} further shows that the spectral shape imposes 
a limit on the diameter and inclination of the model disk, since
any model that causes the disk to cover the bright northern hot spot
would lead to the appearance of a third absorption feature.

The disk models also predict the line of sight velocity toward
the quasar nucleus. This value 
is plotted in each panel of Fig.~\ref{spectra.fig}.  The
average value for the narrow resonance line absorption lies 
between  the components of the double-featured 21cm line profile, 
as indicated in the figure by the point with error bars at the
bottom of each panel.  In this diagram, the error bars actually specify the 
full-width at half-maximum of the resonance line absorption measured by
Foltz, Chaffee \& Wolfe (1988); in order to account for the wide
velocity spread of the metal lines, there needs to be
a ``turbulent medium,'' probably composed of multiple, low column
density clouds spread over more than 100 km~s$^{-1}$. Such a medium
may mask the presence of a
cold, higher column density absorber at the velocity where the
line of sight penetrates the extrapolation of the
intervening galaxy disk.  These separate components
might be deciphered in a high spectral resolution observation of
low oscillator strength metal line transitions, permitting a better
separation of an extended, cold disk component from the turbulent component.

\begin{figure}
  \psfig{figure=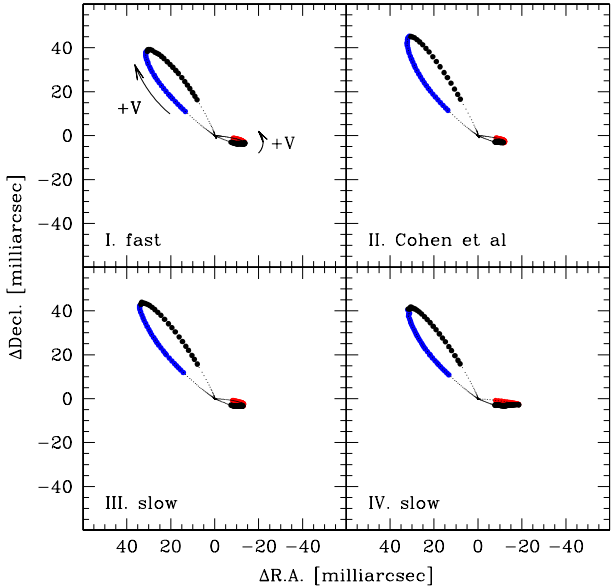,width=8.5cm,angle=0}
  \caption[]{Model calculations of the spectral dependence
of the source centroid for the four representative models. 
The unabsorbed (17.8 Jy) continuum source centroid is located at the
origin $(0,0)$ of the plot. The excursions to the NE (in the 
deeper absorption feature) are given large markers when the 
line becomes deeper than 0.15 Jy; to the East (shallow broad
feature), large markers are used when the depth exceeds 0.04 Jy.
The plots are to be compared with observations in Fig.~\ref{centroid.fig}.
}
  \label{centroids.fig}
\end{figure}

Fig.~\ref{centroids.fig} indicates the location of the source
centroid as a function of frequency through the 21cm line profile
for comparison with Fig.~(3). All models that produce reasonable
``integrated absorption'' profiles also lead to acceptable  trajectories
for the source centroid.

\section{Discussion of Models}

There are several considerations when 
appraising the models for the kinematics of 3C196-G1. The
new 21cm line spectra showing the double-feature profile spread over
$\sim$300 km~s$^{-1}$ places the new constraints on the physical extent
of the absorbing
layer and the kinematics of the gas, and, along with the self-consistent 
centroid shift, these facts confirm  that
3C196-G1 is indeed the cause of the 21cm/damped Lyman-$\alpha$
absorption. It is also of great interest to understand the
nature of the metal-line absorption along the line of sight
to the optical nucleus of 3C196, and a satisfactory interpretation
of the morphology in the HST image would be a useful outcome.
Is it possible to deduce a viewing angle that allows the galaxy
to be deprojected to resemble a normal barred spiral with gently
winding arms, or must the galaxy's morphology be a result of
tidal interaction, possibly caused by 3C196-G2? 
 
We first consider the advantages and disadvantages of each of the
representative kinematic models in turn:

\noindent {\it Model I} does well at reproducing
  the 21cm absorption profiles. This model has a high rotation
  speed $V_{rot}=250$ km~s$^{-1}$, and it is oriented 
  (see Fig.~\ref{velfields.fig}) with the kinematic major axis
  lying closer to the line of sight to the quasar than to the
  hot spot in the SW radio continuum lobe.  This leads to the 
  prediction that the interception velocity towards the QSO nucleus
  will be blue shifted with respect to the deep 21 cm feature,
  in contradiction to what is observed.

\noindent {\it Model II} adopts the orientation and inclination deduced
  for the central stellar body of G1 by Cohen et al (1996). In our
 kinematic model, the radial extent of the disk is increased to cover the
  background radio continuum lobes  with $pa$ and  $i$ fixed at the
  values of Cohen et al. For this model, the interception velocity on
the line of sight to the QSO nucleus falls outside the 21 cm profile,
but not as greatly blue shifted as for Model I.  The predicted optical/UV
velocity falls at the ends of the error bars indicating the measured
UV line velocity spread, implying that high spectral resolution optical
observations could provide a significant test of the viability of this
model.

\noindent {\it Model  III} does nicely on most of the observational tests.
  It produces the correct velocity spread, and the galaxy's optical morphology
  shows winding spiral arms.  Since the line of sight to the quasar optical
  nucleus is closer to the minor axis of the orbits than is the hot spot
  in the SW continuum lobe, the predicted velocity component for the interception 
  of the disk lies on the low redshift side of the deep 21cm feature, instead
  of the high redshift side.
  This may not be a serious problem since the disk interception velocity
  does lie just at one end of the FWHM velocity spread of the metal lines,
  and, in these sorts of MgII/FeII lines, the turbulent width is actually
  built up from many narrower components that represent individual absorbing
  clouds. Furthermore, 
  it is expected that point-like interceptions of differentially
  rotating disks that are either thick or located within a corotating
  halo of cloudlets will produce asymmetric metal-line profiles -- with the
  21cm line or thickest, cold component at one side of the profile
 (Briggs et al 1985, Lanzetta \& Bowen 1992, Prochaska \& Wolfe 1997, 1998).

\noindent {\it Model IV}  is minor re-orientation of model III with the 
  goal of increasing the degree of winding of the arms in the deprojected
  image and decreasing the radial extent of the gas 
  (see Fig.~\ref{deprojoverlays.fig}). This relatively small change from
  model III produces distortion in the low redshift 21cm feature due to
  rotation of the major axis of the rotating disk onto the more intense
  part of the diffuse eastern lobe.  The narrow feature that develops is 
  the beginning of a ``horn'' of a conventional double-horned emission
  profile observed in $z=0$ disk galaxies.  This model IV has a near
  perfect agreement between the observed metal-line redshift and
  the disk interception velocity.

We have adopted models I and III
for working hypotheses in the subsequent discussions, since they
represent the extremes in rotation velocity while still yielding
reasonable velocity profiles. However, there clearly
are variations of these models, such as non-planar and non-circular
motions (warps, tidally induced arms, etc) and non-uniform gas coverage
that would add many additional degrees of freedom to any model that strives to
fit the absorption profiles exactly.  The strongest constraints are
that the rotation velocity must be at least ${\sim}150$~km~s$^{-1}$ and that
the velocity field is centered on the galaxy G1. Larger rotation speeds
can always be compensated to some extent
by lowering the inclination $i$ and adjusting
the optical depth coverage to reproduce the observed profiles.  
However, the inclination cannot be lowered much from the tabulated values
before the bright continuum hot-spot in the NE lobe becomes covered, 
which would add an additional strong absorption component to the model
profiles.  The uniform disk models I and III would reach the  NE hot spot
when $i$ is reduced from 70$^{\circ}$ to 65$^{\circ}$ and 50$^{\circ}$,
respectively.  For these models to still provide the observed
velocity spread as the inclination is lowered,
the disk rotation speed would have to increase by 4\% for model I and
22\% for model III.

The sharp edge at $R_{\rm 21}$
assumed in the  simple uniform disk model is unlikely to
be realistic, and these data are unable to address the question of
whether there is a hole in the neutral gas at the center of the
galaxy where the bright stellar body is located.
One of the most viable alternatives to the simple uniform disk model
involves a  tidal interaction, which would give rise
to long gas-rich arms that could extend nearly radially from the host
galaxy and could be highly non-planar.  One possibility is that the
galaxy G2 (located slightly to the north and east of the quasar nucleus)
is interacting with G1, although other investigations favor the association
of G2 with the quasar at $z_{em}=0.871$ (Ridgway \& Stockton 1997).

All of the disk models in Table 1 are consistent with Brown etal (1988)
988 MHz spectral-line
VLBI result, which found no trace of 21cm absorption against the
northeastern radio-lobe hot spot, since all the models in
Table~\ref{model.tab} cut off
at radii that do not reach the hot spot. A tidal arm model would 
similarly place no absorbing gas over the hot spot. However, if
the second galaxy G2 is actually a perturber of G1, it may be
useful to observe a still broader spectrum in the 21cm line to
test whether G2 has a low-level gaseous envelope that extends to
the hot spot.

The line of sight column density of HI is $2.1{\times}10^{20}$cm$^{-2}$ for
model II and $5.6{\times}10^{20}$cm$^{-2}$ for models I, III and IV.
Spin temperatures in redshifted absorption systems are generally higher than
the $T_s = 100$~K that we have adopted here (Wolfe \& Davis 1979,
Wolfe et al 1985, Taramopoulos et al 1995, Carilli et al 1996,
Briggs et al 1997, Chengalur \& Kanekar 2000), implying that the
$N_{HI}$ may be considerably higher and that this system definitely
lies in the DLa class.
Cohen et al (1996) measured the Lyman-$\alpha$ line in an
HST spectrum, but they discovered that this region of the
spectrum was confused by the higher order Lyman-series lines
of a higher redshift Lyman-limit system at ${\sim}z_{em}$ and metal
absorption lines.  Their analysis is compatible with a range
for $N_{los}(HI)$ of  $2.7{\times}10^{19}$ to $1.5{\times}10^{20}$cm$^{-2}$,
with a best-fit favoring the highest value.  Thus, the
preferred value of Cohen et al
is in reasonable agreement with the range of column
densities required in our models. The agreement provides evidence that 
the HI in the galaxy is distributed throughout a disk, and not
concentrated solely in the strong arms that must dominate in the
radio absorption spectrum, since the HST image shows the arms covering
the significant areas of the radio lobes.

In the future, 
the models could be further constrained by synthesis mapping of the 21cm
absorption
system at ${\sim}0.5''$ resolution.  There is significant low
level extended, steep spectrum radio emission that would allow a
sensitive observation to map absorption by 
a substantial portion of the outlying gas.

\begin{figure}
 \psfig{figure=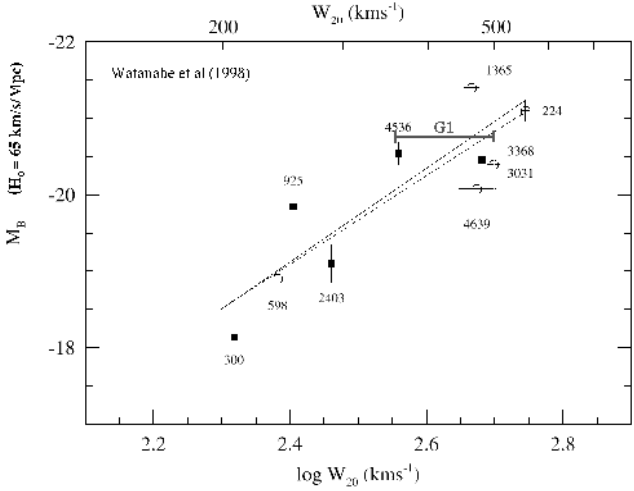,width=8.5cm,angle=0}
  \caption[]{Tully-Fisher Relation. Absolute B-band magnitude
is plotted as a function of galaxy velocity width  (here taken to
be $W=2V_{rot}$).
An error bar representing galaxy
3C196-G1 is plotted on the graph of Watanabe et al (1998). Length of
the bar represents the range of $V_{rot}$ for the models I-IV. Absolute
magnitude for G1 is the rest frame B-band value from Cohen et al (1996)
corrected to a Hubble constant of 65~km~s$^{-1}$~Mpc$^{-1}$.
}
  \label{TF.fig}
\end{figure}

The comparison of dynamical mass to luminosity  for the galaxy 3C196-G1
is straightforward.  Cohen et al (1996) derived a rest-frame B-band
absolute magnitude of ${\rm M}_B$ of $-21.3$ with Hubble Constant of
$H_o = $50~km~s$^{-1}$Mpc$^{-1}$. This point, adjusted to 
$H_o = $65~km~s$^{-1}$Mpc$^{-1}$ is plotted on the Tully-Fisher
Relation of Watanabe et al (1998) in Fig.~\ref{TF.fig} using the
spread in $V_{rot}$ from 180 to 250~km~s$^{-1}$. The range as plotted
is  consistent with the scatter of the local Tully-Fisher
calibrators, although no corrections for extinction
have been applied to the ${\rm M}_B$ measurement, making it a lower limit to
optical luminosity. 

\section{Conclusion}

 The location of the absorbing gas leaves no doubt that the
barred spiral seen in the HST image is responsible for the 21cm
absorption. 
The western hotspot and eastern lobe of 3C196 are
located exactly on top of spiral arm-like features emanating from
the main body of the galaxy. The total velocity spread is about 250 km/sec,
consistent with that expected from a massive spiral galaxy at
moderate inclination. The implied column density of HI gas, for
an adopted spin temperature of the HI of 100 K, is then about $10^{21}$
atoms~cm$^{-2}$. This shows that we are dealing with a very
gas rich spiral with a projected dimension of at least 6$''$ or
40 kpc and a total amount of $10^{10}M_{\odot}$ of
HI. 

 This result shows that HI absorption study of extended,
cosmologically distant, radio sources is an excellent method to
probe the spatial distribution and kinematics of disks of HI at
high redshifts. The median size of high redshift radio galaxies
(about 50 kpc) is ideally matched to the size of gaseous disks.
However, to apply this technique to more radio sources one would
like to have about 30 times better angular resolution since not
all radio sources are as bright as 3C196 and the use of source
centroid displacement as a function of frequency as a `mapping'
technique is only viable  in cases of exceptionally
`deep' absorption lines (deep in terms of the noise level).
Searches for such`damped Ly-alpha systems' in high redshift
radio sources are currently underway.

\begin{acknowledgements} 
  The authors are grateful to the staff of the Westerbork Telescope
  for their skill and dedication in bringing the wide-band UHF receiving
  systems into operation.  We are grateful to S.E. Ridgway 
 for kindly providing us with a FITS file of her fully processed
  HST image.
The Westerbork Synthesis Radio Telescope (WSRT) is operated by the
Netherlands Foundation for Research in Astronomy (NFRA) with financial
support of the Netherlands Organization for Scientific Research (NWO).
This research has made use of the NASA Astrophysics Data
  System (ADS) and the NASA/IPAC Extragalactic Database
  (NED) which is operated by the Jet Propulsion Laboratory, California
  Institute of Technology, under contract with the National
  Aeronautics and Space Administration. 

\end{acknowledgements}

\end{document}